\documentclass[sigconf]{acmart}

\AtBeginDocument{%
  \providecommand\BibTeX{{%
    Bib\TeX}}}

\acmDOI{}

\acmConference[SIGSPATIAL]{}{June 2023}{}
\acmPrice{15.00}
\acmISBN{978-1-4503-XXXX-X/18/06}

\usepackage[ruled,vlined,linesnumbered]{algorithm2e}
\SetCommentSty{small}
\usepackage{tikz}
\usepackage{pgf}
\usepackage{multirow,tabularx}
\usepackage{booktabs}
\usepackage{subfigure}
\usepackage{xcolor,colortbl}
\usepackage{threeparttable}
\definecolor{lavender}{rgb}{0.9, 0.9, 0.98}
\definecolor{asparagus}{rgb}{0.53, 0.66, 0.42}
\definecolor{caribbeangreen}{rgb}{0.0, 0.8, 0.6}
\definecolor{darkolivegreen}{rgb}{0.33, 0.42, 0.18}
\definecolor{darkpastelgreen}{rgb}{0.01, 0.75, 0.24}
\begin{document}
\title{Popularity-based Alternative Routing}


\author{Giuliano Cornacchia}
\affiliation{
   \institution{ISTI-CNR \\University of Pisa}
\city{Pisa}
\country{Italy}} \email{giuliano.cornacchia@phd.unipi.it}

\author{Ludovico Lemma}
\affiliation{
   \institution{University of Pisa}
\city{Pisa}
\country{Italy}
}
\email{ludovicolemma@gmail.com}

\author{Luca Pappalardo}
\affiliation{
\institution{ISTI-CNR \\ Scuola Normale Superiore}
\city{Pisa}
\country{Italy}
}
\email{luca.pappalardo@isti.cnr.it}

\renewcommand{\shortauthors}{}

\begin{abstract}

Alternative routing is crucial to minimize the environmental impact of urban transportation while enhancing road network efficiency and reducing traffic congestion. Existing methods neglect information about road popularity, possibly leading to unintended consequences such as increasing emissions and congestion. 
This paper introduces {\scshape{Polaris}}, an alternative routing algorithm that exploits road popularity to optimize traffic distribution and reduce CO2 emissions. {\scshape{Polaris}} leverages the novel concept of $K_{\text{road}}$ layers, which mitigates the feedback loop effect where redirecting vehicles to less popular roads could increase their popularity in the future. We conduct experiments in three cities to evaluate {\scshape{Polaris}} against state-of-the-art alternative routing algorithms. 
Our results demonstrate that {\scshape{Polaris}} significantly reduces the overuse of highly popular road edges and traversed regulated intersections, showcasing its ability to generate efficient routes and distribute traffic more evenly.
Furthermore, {\scshape{Polaris}} achieves substantial CO2 reductions, outperforming existing alternative routing strategies. Finally, we compare {\scshape{Polaris}} to an algorithm that coordinates vehicles centrally to distribute them more evenly on the road network. Our findings reveal that {\scshape{Polaris}} performs comparably well, even with much less information, highlighting its potential as an efficient and sustainable solution for urban traffic management.
\end{abstract}




\maketitle

\section{Introduction}
\label{sec:introduction}

Alternative routing is crucial in real-world scenarios where recommending only the shortest path is insufficient \cite{abraham2013alternative, luxen2012candidate}. These scenarios include navigation services offering longer but more desirable routes (e.g., green, scenic, safe), transporting humanitarian aid through dangerous regions by using non-overlapping routes to ensure delivery, and in emergencies like natural disasters or terrorist attacks to prevent panic and collisions.

One way to find alternative routes is by providing a set of disjointed routes \cite{suurballe1974disjoint}. However, this approach typically results in routes that are much longer than the shortest path, making them less practical for drivers. Other methods focus on solving the $k$-shortest paths problem without the requirement of disjointness \cite{yen1971finding, aljazzar2011k, cheng2019shortest}. The alternative routes these methods generate often overlap significantly ($\gg$ 90\%), resulting in slight variations of the same route.
Another solution addresses this issue by finding the $k$-shortest path with limited overlap, offering alternative routes that overlap less than a certain extent \cite{suurballe1974disjoint}. 
A notable example is $k$-Most Diverse Near-Shortest Paths (KMD), which optimises route diversity while still ensuring that the alternative routes do not exceed a user-defined length limit \cite{hacker2021most}. 

\begin{figure}
    \centering
    \includegraphics[width=0.9\columnwidth]{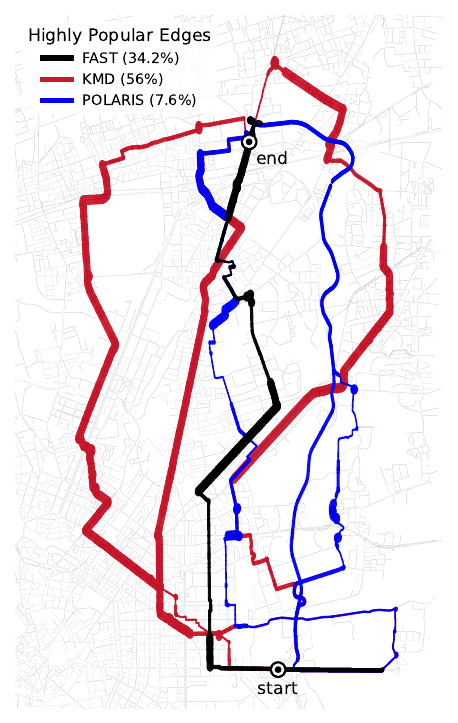}
    \caption{Comparison of three alternative routes generated by KMD (in red), three alternative routes generated by {\scshape Polaris} (in blue), and the fastest path (in black) between two roads in Milan. The width of the roads is proportional to their popularity, computed as their $K_{\text{road}}$. KMD routes differ significantly from the fastest path but use much more popular roads (56\%). In contrast, {\scshape Polaris} minimizes the use of popular road edges (only 7.6\%), while still ensuring diverse route options.}
    \label{fig:example}
\end{figure}

All these solutions have a major limitation: they do not consider the \emph{road popularity}, i.e., how many areas of the city are routing vehicles through a road, indicating the potential congestion on that particular road. 
In real-world scenarios where alternative routing is implemented in digital systems, neglecting information about popularity can lead to unintended consequences, such as increasing emissions and congestion and leading to an inefficient use of the road network \cite{pappalardo2023future, navigation, pedreschi2024humanai}.
Figure \ref{fig:example} helps clarify this concept. 
It displays three alternative routes provided by KMD (highlighted in red) between two roads. 
These three routes are quite diverse from the shortest path (black line), but primarily use popular road edges. This suggests that KMD, the state-of-the-art algorithm for alternative routing, may create bottlenecks in the road network.
In the same figure, we present in blue the alternative routes provided by {\scshape Polaris}, the algorithm we introduce in this paper. {\scshape Polaris} minimises the use of popular roads while still ensuring diverse route options.

{\scshape{Polaris}} is designed to adjust road edge weights by considering road popularity. This strategy aims to alleviate traffic congestion by avoiding the excessive use of highly popular roads. Our approach leverages the concept of $K_{\text{road}}$ layers, which assists in counteracting a feedback loop effect: directing traffic towards lesser-used roads may inadvertently boost their popularity over time.

Our experiments in three cities compare {\scshape{Polaris}} with various state-of-the-art alternative routing approaches. 
{\scshape{Polaris}} significantly reduces the use of highly popular roads compared to the baselines. This outcome translates into a lighter environmental footprint in terms of CO2 emissions, with reductions of up to 23.57\% compared to the best baselines. 
We also compare {\scshape{Polaris}} against an approach that coordinates vehicles to distribute traffic more evenly on the road network. Our algorithm performs comparably to the coordinated approach, even exploiting less traffic information.

In summary, this paper provides the following key contributions:
\begin{itemize}
\item We introduce the concept of $K_{\text{road}}$ layers, which consider the popularity of road edges while considering the potential for redirecting vehicles to unpopular roads to increase their future popularity;
\item We develop am edge weight penalization strategy based on $K_{\text{road}}$ layers to generate alternative routes that minimize the use of popular roads; 
\item We conduct extensive experiments to demonstrate the superior performance of {\scshape{Polaris}} over existing alternative routing algorithms in reducing CO2 emissions while maintaining competitive computational performance.
\end{itemize}

\subsubsection*{\bf Open Source}
The code that implements {\scshape Polaris} and the baselines is available at \url{https://pypi.org/project/pattern-optimized-routes/}.

\section{Related Work}
\label{RW}

The fastest route is the most direct way for connecting two points within a road network \cite{wu2012shortest}.
The $k$-shortest path approaches seek to identify the $k$ shortest routes between an origin and a destination \cite{yen1971finding, aljazzar2011k}. 
In practical settings, these solutions often lack route diversification, with routes overlapping by 99\% in terms of road edges \cite{cheng2019shortest}.
The $k$-shortest disjointed paths approach \cite{suurballe1974disjoint} aims to find $k$ routes that significantly deviate from the shortest path but do not overlap, leading to notable increases in travel time. 
Several approaches exist between the $k$-shortest path and $k$-shortest disjoint paths paradigms, which can be categorized into edge-weight, plateau, and dissimilarity approaches.

\emph{Edge-weight approaches.} These methods compute the shortest routes iteratively. During each iteration, the edge weights of the road network are updated to compute $k$ alternative routes. This updating process may include randomizing the weights or applying cumulative penalties to edges that compose the shortest routes. While edge-weight approaches are easy to implement, they do not guarantee the generation of significantly diverse routes \cite{li2022comparing}.

\emph{Plateau approaches.} 
These methods build two shortest-path trees, originating from the source and destination nodes. 
They identify common branches between these trees, referred to as plateaus \cite{camvit2005choice}. The top-$k$ plateaus are selected based on their lengths, and alternative routes are derived by appending the shortest paths from the source to the first edge of the plateau and from the last edge to the target. 
Due to their disjointed nature, plateaus may result in considerably longer routes than the fastest path \cite{camvit2005choice}.

\emph{Dissimilarity approaches.} 
These methods generate $k$ paths that adhere to a dissimilarity constraint and a specified criterion. 
\citet{liu2018finding} introduce $k$-Shortest Paths with Diversity ($k$SPD), which identifies the top-$k$ shortest paths that are maximally dissimilar while minimizing their total length. 
\citet{chondrogiannis2015kshortest} propose an implementation of the $k$-Shortest Paths with Limited Overlap ($k$SPLO) to recommend $k$ alternative routes that are as short as possible while maintaining sufficient dissimilarity. Additionally, \citet{chondrogiannis2018mincoll} formalize the $k$-Dissimilar Paths with Minimum Collective Length ($k$DPML) problem, wherein a set of $k$ routes containing adequately dissimilar routes and the lowest collective path length is computed given two road edges.  
\citet{hacker2021most} introduce $k$-Most Diverse Near Shortest Paths (KMD) to recommend the set of $k$ near-shortest routes with the highest diversity, based on a user-defined cost threshold. 
However, dissimilarity approaches do not guarantee the existence of a set of $k$ routes satisfying the desired property.
\citet{cornacchia2023} propose {\scshape{Metis}}, a traffic assignment algorithm that incorporates vehicle coordination, alternative routing and edge-weight penalization to diversify routes on the road network.

\section{{\scshape{Polaris}}} \label{ERU}

We propose an edge-weight penalization approach called {\scshape{Polaris}} (POpularity-based aLternAtive RoutIng Strategy) that adjusts edge weights to prevent the overuse of highly popular roads. 
{\scshape{Polaris}} incorporates the principle of $K_{\text{road}}$ layers to minimize the effect of a feedback loop effect: rerouting vehicles to unpopular roads could potentially increase their popularity in the future.

\subsection{$K_{\text{road}}$}
\label{sec:kroad}
The concept of $K_{\text{road}}$ layer is based on measuring $K_{\text{road}}$, which proxies road popularity by quantifying how many areas of the city contribute to most of the traffic flow over a road edge \cite{wang2012understanding, baccile2024measuring, cornacchia2023}.

$K_{\text{road}}$ is computed over a road usage network, which is a bipartite network where each road edge is linked to its major driver areas, i.e., those areas responsible for 80\% of the traffic flow on that edge \cite{wang2012understanding}. The value $K_{\text{road}}(e)$ for a road edge $e$ indicates the network degree of $e$ within the road usage network. A low $K_{\text{road}}$ indicates that the edge is used by only a limited number of sources, making it relatively unpopular. Conversely, a high $K_{\text{road}}$ indicates that the road edge attracts traffic from many areas, making it more popular.

\begin{figure*}
    \centering
    \subfigure[Florence]{
    \includegraphics[width=0.38\textwidth]{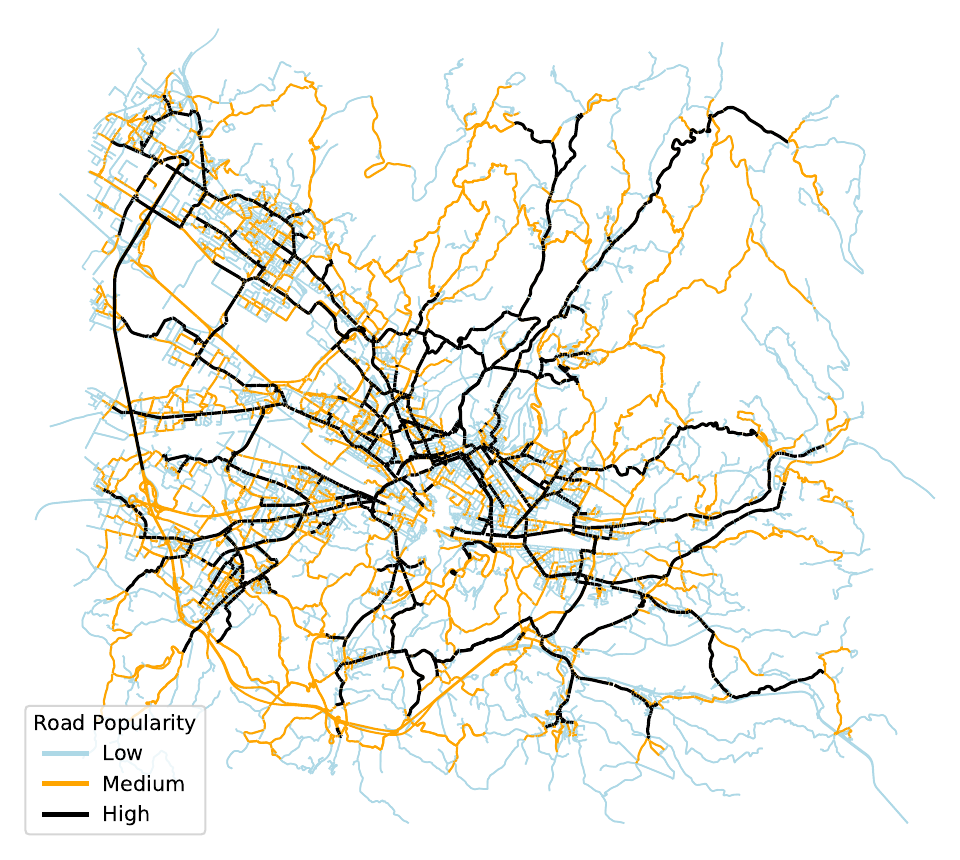}}
    \subfigure[Milan]{\includegraphics[width=0.27\textwidth]{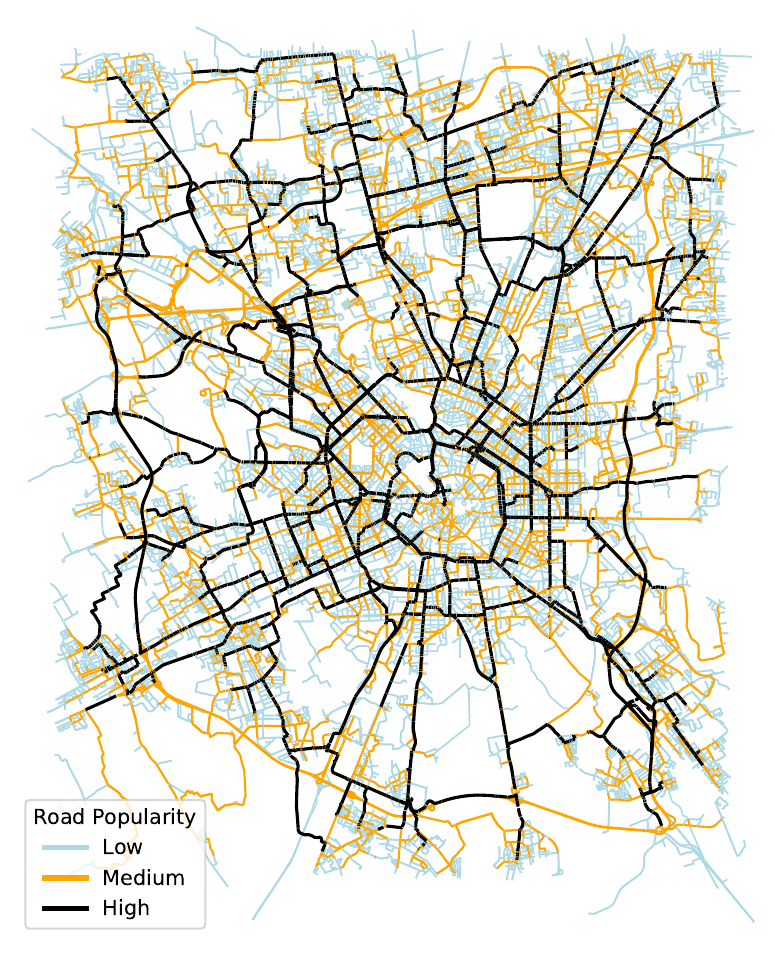}}
    \subfigure[Rome]{
    \includegraphics[width=0.33\textwidth]{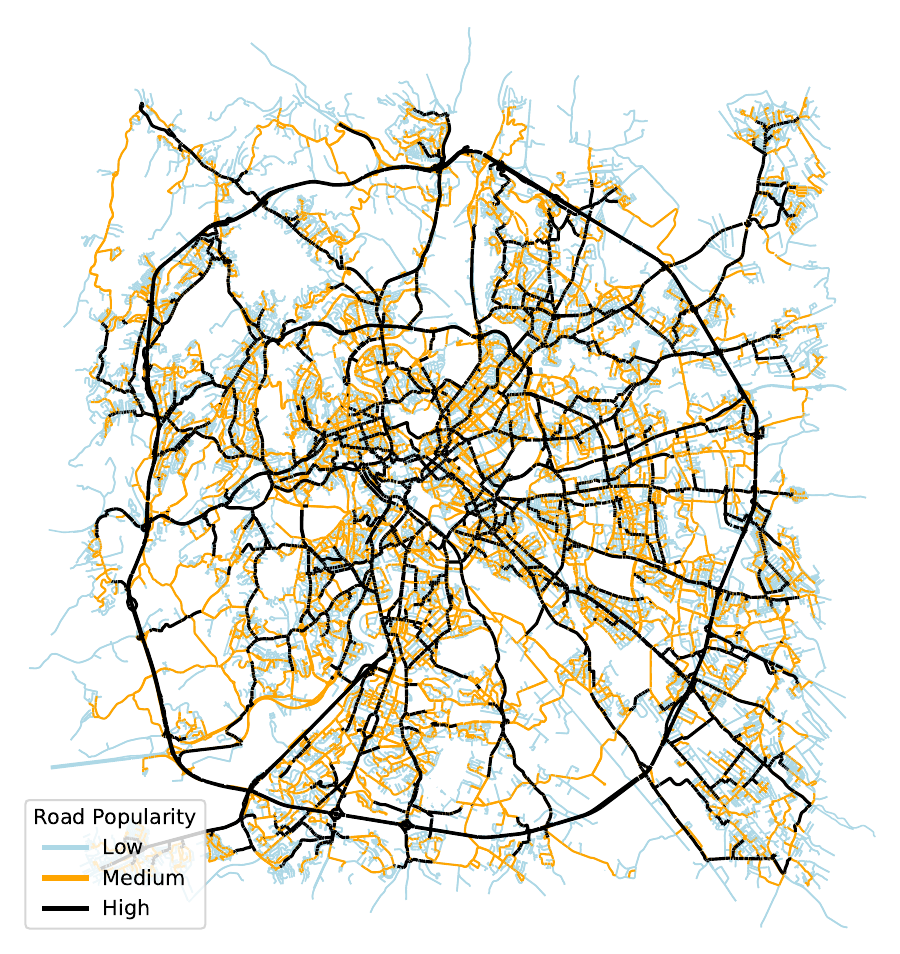}}
    \caption{Spatial distribution of road popularity, computed as $K_{\text{road}}$, in Florence (a), Milan (b), and Rome (c). We categorise road edges into Low (cyan), Medium (orange) and High (black) popularity using an equal-sized logarithmic binning on the $K_{\text{road}}$ distribution of each city.
    Note how, in all cities, there are many highly popular road edges (black road edges).} 
    \label{fig:maps_kroad}
\end{figure*}

To compute $K_{\text{road}}$, we need to gather a set of routes to estimate the sources and destinations of traffic on the road network. In contrast to \citet{wang2012understanding}, who use real GPS data to compute $K_{\text{road}}$ for each road edge, we adopt a more adaptable strategy \cite{baccile2024measuring, cornacchia2023}. We randomly select $v$ origin-destination pairs on the road network and connect them using the fastest route, assuming free-flow travel time. We then use these routes to compute $K_{\text{road}}$ for each road edge in the network. 
We normalize $K_{\text{road}}$ to fall within the range $[0, 1]$ using min-max normalization.
Figure \ref{fig:maps_kroad} displays the $K_{\text{road}}$ value of roads in Florence, Milan, and Rome. 
The figure reveals, in all cities, the presence of many highly popular road edges.

\subsection{$K_{road}$ layers}
\label{sec:kroad_layers}
By penalizing road edges with a high $K_{\text{road}}$, we direct traffic onto less popular edges. 
This creates a feedback loop process: the originally less-used roads become popular due to the algorithm's recommendations, while the originally popular roads become underused.

To address this feedback loop, we propose a dynamic approach in which $K_{\text{road}}$ values are iteratively updated. This strategy involves computing multiple $K_{\text{road}}$ values for each edge in the road network, referred to as $K_{\text{road}}$ \emph{layers}.  

Algorithm \ref{getkroadlevels} describes the procedure for computing these $K_{\text{road}}$ layers. 
The inputs of the algorithm are the road network, represented as a directed weighted graph $G = (V, E)$ where $V$ is the set of intersections and $E$ is the set of road edges; the number $v$ of trips used to estimate $K_{\text{road}}$; and the number $m$ of $K_{\text{road}}$ layers. 

For each of the $m$ layers (lines 2-12), we randomly sample $v$ origin-destination nodes within $G$ and compute the fastest routes between them (lines 4-7). Based on these $v$ fastest routes, we calculate $K_{\text{road}}$ for each road edge in $G$ (line 8) and normalize these values to the range $[0,1]$ (line 9). 
In line 10, the normalized $K_{\text{road}}$ for each edge is assigned to the current $K_{\text{road}}$ layer. 
Following this, edge weights in the road network are penalized according to the current $K_{\text{road}}$ layers (lines 11-12). 
This process repeats until all $m$ $K_{\text{road}}$ layers are computed. 

At the end of the procedure, each edge $e \in E$ is associated with a list of $m$ values: 
$$
\mathcal{K}(e) = [K_{\text{ road}}^{(1)}, \dots, K_{\text{ road}}^{(m)}]
$$
The computational cost of this procedure is $\mathcal{O}(m \cdot n \cdot ((|V|+|E|) \cdot log_2 |V|))$, and it only needs to be performed once.

\begin{algorithm}
    \SetKwInOut{Input}{Input}
    \SetKwInOut{Output}{Output}
    
    \Input{Graph $(V, E)$: $G$; weight of edge $e \in E$: $w_e \in w$; number of OD pairs $v$; number of layers: $m$}
    \Output{List of K$_{\text{road}}$ values for each edge $e \in E$}

      $L \leftarrow []$\ ;
      
      \For{$l = 0$ \textbf{to} $m-1$} {
         $P \leftarrow \emptyset$\ ; 
        
        \While{$|P| < v$} {
          
          

           $(o, d) \leftarrow \text{get\_random\_od}(G)$ 
          
           $p \leftarrow \text{get\_shortest\_path}(G, o, d, w)$\ ;
          
           $P \leftarrow P \cup \{p\}$ ;
          
        }

         $S \leftarrow$ compute\_kroad($P$) ;
        
         $\mathcal{K}_l \leftarrow$ min\_max\_normalization($S$) ;
        
         $L[l] \leftarrow \mathcal{K}_l$ ;
        
        \For{$e \in E$} {
           $w_e \leftarrow w_e \cdot (1 + \mathcal{K}_l[e])$ ;
        }
      }
      
      \Return $L$ ;
\caption{Computation of $K_{road}$ layers.}
\label{getkroadlevels}
\end{algorithm}

\subsection{Multi-layer Edge-Weight Penalization}
\label{sec:meru}
Algorithm \ref{multilevelapproach} provides a high-level pseudocode for {\scshape{Polaris}}. It requires three inputs: the road network $G$ with each edge associated with its $K_{\text{road}}$ layers and expected travel time, estimated by dividing its length by the maximum speed allowed; the origin and destination pair, for which we must compute the alternative routes; the number $k$ of desired alternative routes.

During each iteration $i$ (lines 3-11), {\scshape{Polaris}} updates the cost of each edge in the road network by multiplying it with the $K_{\text{road}}$ value in the $i$-th layer or the last $K_{\text{road}}$ layer if $i \geq m$ (lines 4-6). The algorithm then computes the shortest route (line 7) on the adjusted road network and re-penalizes edge weights in the current shortest routes proportionally to their $K_{\text{road}}$ in layer $0$ (i.e., the original $K_{\text{road}}$ value) (line 9-10). This process helps ensure that subsequent shortest routes differ from the previous ones, preventing certain edges from being overused. This iterative process continues until the desired number $k$ of routes is obtained.

\begin{algorithm}
\caption{{\scshape{Polaris}}}
\label{multilevelapproach}
    \SetKwInOut{Input}{Input}
    \SetKwInOut{Output}{Output}
    
    \Input{Graph $(V, E)$: $G$; weight of edge $e \in E$: $w_e \in w$; Set of $m$ k-road levels $L = \{\mathcal{K}_0, \ldots, \mathcal{K}_{m-1}\}$, with $K_{road}$ of level $l$ for edge $e$ defined as: $\mathcal{K}_l[e] \in \mathcal{K}_l$; origin vertex: $o$; destination vertex: $d$; number of alternative paths: $k$}
    \Output{Set of $k$ paths $P = \{p_1, p_2, \ldots, p_k\}$}

      $P \leftarrow \emptyset$\ ;
      
      $i \leftarrow 0$ ;

      \While{$|P| < k$} {
        $\mathcal{K}_i \leftarrow L[min(i, m-1)]$ ;
        
        \For{$e \in E$} {
          $w_e \leftarrow w_e \cdot (1 + \mathcal{K}_i[e])$\ ; }

        $p \leftarrow \text{get\_shortest\_path}(G, o, d, w)$\ ;
        
        $P \leftarrow P \cup \{p\}$\ ;

        \For{$e \in p$} {
          $w_e \leftarrow w_e \cdot (1 + \mathcal{K}_0[e])$\ ; }

        $i \leftarrow i + 1$ ; 
        }

      \Return $P$\ ;
\end{algorithm}

\section{Experimental Setup} \label{experiments}
This section describes the experimental settings (Sections \ref{sec:experimental_settings}), the baselines we compare with {\scshape{Polaris}} (Section \ref{baselines}), and the measures used for this comparison (Section \ref{sec:measures}).

\subsection{Experimental settings}
\label{sec:experimental_settings}
We conduct experiments in three Italian cities: Milan, Rome, and Florence. These cities represent diverse urban environments with varying traffic dynamics, sizes, and road networks (see Table \ref{tab:road_networks}).

\subsubsection*{\bf Road Networks}
We get the road network for each city using OSMWebWizard.\footnote{\url{https://sumo.dlr.de/docs/Tutorials/OSMWebWizard.html}} The road network characteristics of the three cities are heterogeneous (see Table \ref{tab:road_networks}). Florence, the smallest city, has the second-highest road network density. In contrast, Milan and Rome have more extensive road networks. Among these cities, Rome has the densest road network. This variation in road network characteristics allows us to evaluate the performance of alternative routing algorithms in different urban contexts.

\subsubsection*{\bf Mobility Demand}
\label{sec:mobility_demand}
We split each city into 1km squared tiles using GPS traces \cite{scikitmob} to determine each vehicle's trip's starting and ending tiles. 
We use this information to create an origin-destination matrix $M$, where $m_{o, d}$ represents the number of trips starting in tile $o$ and ending in tile $d$.
To generate a mobility demand $D$ of $N$ trips, we randomly select trips $T_v=(e_o, e_d)$ choosing matrix elements with probabilities $p_{o, d} \propto m_{o, d}$. 
We then uniformly select two edges $e_o$ and $e_d$ within tiles $o$ and $d$ from the road network $G$.

\begin{figure}
    \centering
    \includegraphics[width=1\columnwidth]{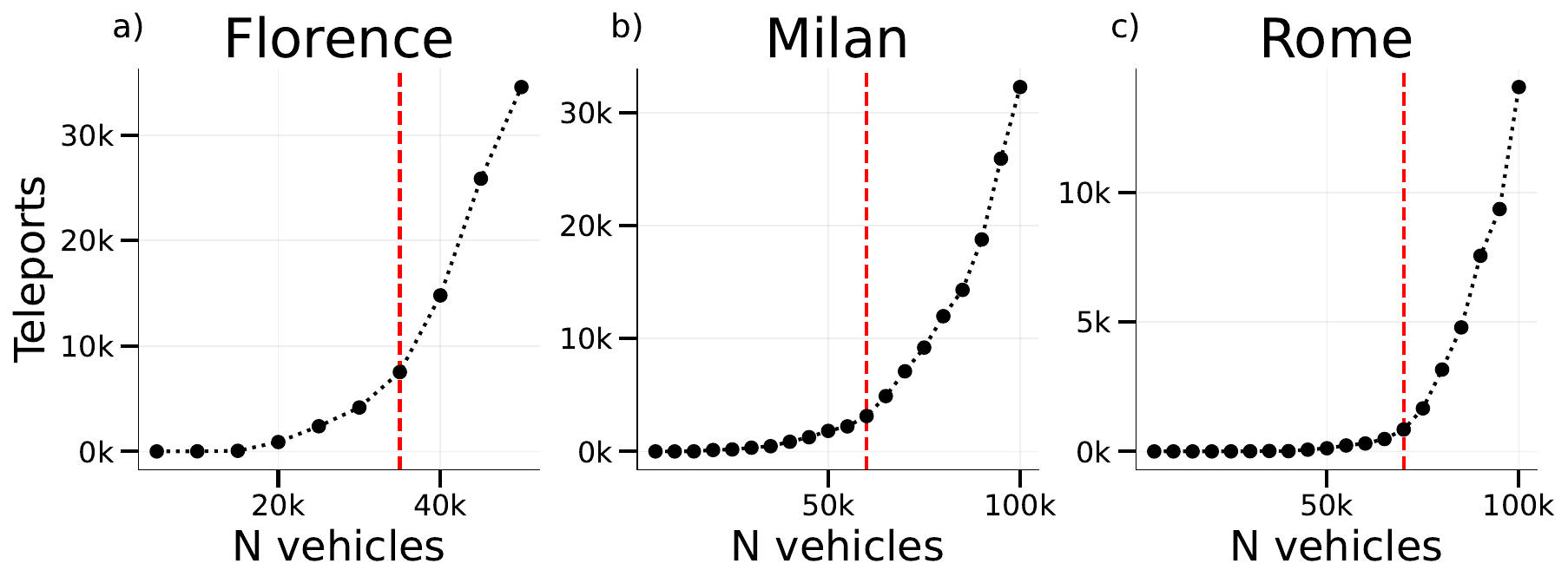}
    \caption{Number of vehicles in the simulation versus the number of SUMO teleports for Florence, Milan, and Rome. Simulations are conducted based on the mobility demand of the city and using the fastest route.
    The red dashed line denotes the elbow point of the curve.}
    \label{fig:teleports}
\end{figure}

In our experiments, we set $N$=35k trips in Florence, $N$=60k in Milan, and $N$=70k trips in Rome. To select these values, we run simulations for each city and algorithm, using SUMO (Simulation of Urban MObility) \cite{Microscopic2018Lopez, cornacchia2022how, cornacchia2023}, a traffic simulator that models vehicle dynamics, interactions and traffic congestion. 
In each simulation, we generate the fastest route for each trip in the city's mobility demand and provide this set of routes to SUMO.
In order to maintain smooth simulations and avoid gridlock, SUMO uses "teleports" to instantly relocate a vehicle stuck in traffic for a long time to the next available road edge of its route. The number of teleports during the simulation reflects its quality: more teleports indicate lower reliability. 
We analyse the relationship between the number of vehicles simulated in the city and the number of teleports, choosing $N$ as the point where the rate of increase in teleports sharply rises, indicating a sharp change in traffic congestion (see Figure \ref{fig:teleports}). 

\begin{table}[htb]
\centering
\begin{tabular}{@{}lrrr|cc}
\toprule
\bf city & \bf $|V|$ & \bf $|E|$ &  $\frac{|E|}{|V|}$ &\textbf{\begin{tabular}[c]{@{}c@{}}\small Total Edge\\\small Length (km)\end{tabular}} & \bf \textbf{\begin{tabular}[c]{@{}c@{}}\small Total Lane\\\small Length (km)\end{tabular}} \\ 
\midrule
Florence &  11,454&  22,728&  1.984& 2,876 &  3,264  \\
Milan & 29,134 &  56,521&  1.940&  5,260&   6,324 \\
Rome &  32,815&  65,352&  1.992&  6,788&   8,301 \\ \bottomrule
\end{tabular}
\caption{Road network characteristics for the three cities. The columns show the number of vertices $|V|$ and edges $|E|$, their ratio $\frac{|E|}{|V|}$, the total edge length (in km), and the total lane length (in km).}
\label{tab:road_networks}
\end{table}

\subsection{Baselines} \label{baselines}

We compare {\scshape{Polaris}} against several alternative routing approaches. 
We exclude iterative solutions such as User Equilibrium (UE) \cite{wardrop1952some, friesz2010dynamic} and System Optimum \cite{wardrop1952proceedings} due to their computational intensity and requirement for multiple iterations, rendering them unsuitable for real-time applications. 
Additionally, we exclude navigation services (e.g., Google Maps and TomTom) from our evaluation. Direct comparison with our proposal under identical traffic conditions is unfeasible because navigation services APIs recommend routes based on real-time and real-world traffic conditions, which may differ from those simulated within our study.

In this study, we consider the following state-of-the-art alternative routing algorithms:

\begin{itemize}
    \item {\bf FAST (Fastest Path)} \cite{traffic-assignment-methods} always assigns the fastest route to connect each trip's origin-destination within a mobility demand;
    
        \item \textbf{PP (Path Penalization)} generates $k$ alternative routes by penalizing the weights of edges contributing to the fastest path \cite{cheng2019shortest}. In each iteration, PP computes the fastest route and increases the weights of the edges that contributed to it by a factor $p$ as $w(e) = w(e)\cdot(1+p)$. The penalization is cumulative: if an edge has already been penalized in a previous iteration, its weight will be further increased \cite{cheng2019shortest};

    \item \textbf{GR (Graph Randomization)} generates $k$ alternative routes by randomizing the weights of all edges in the road network before each fastest route computation. The randomization is done by adding a value from a normal distribution, given by the equation $N(0, w(e)^2\cdot \delta^2)$ \cite{cheng2019shortest};

    \item \textbf{PR (Path Randomization)} generates $k$ alternative routes randomizing only the weights of the edges that were part of the previously computed route.
    Similar to GR, it adds a value from a normal distribution to the edge weights, following the equation $N(0, w(e)^2\cdot \delta^2)$ \cite{cheng2019shortest}.

    \item \textbf{KMD ($k$-Most Diverse Near Shortest Paths)} generates $k$ alternative routes with the highest dissimilarity among each other while adhering to a user-defined cost threshold $\epsilon$ \cite{hacker2021most}.

\end{itemize}

Table \ref{tab:best_params} shows the parameter ranges tested for each baseline and the best parameter combinations obtained in our experiments.

\begin{table}[htb]
\centering
\begin{threeparttable}
\begin{tabular}{@{}lllll@{}}
\toprule
 & & \multicolumn{3}{c}{\bf best params} \\
 \cline{3-5}
\bf algo & \bf params range & \bf Florence & \bf Milan & \bf Rome \\ \midrule
PP & \footnotesize{$p \in \{.1, .2, \dots, .5\}$} & $p=.4$ & $p=.1$ & $p=.1$ \\
PR & \footnotesize{$\Delta \in \{.2, .3, .4, .5\}$} & $\Delta=.2$ & $\Delta=.2$ & $\Delta=.2$ \\
GR & \footnotesize{$\delta \in \{.2, .3, .4, .5\}$} & $\Delta=.2$ & $\Delta=.2$ & $\Delta=.2$ \\
KMD & \footnotesize{$\epsilon \in \{.01, .05, .1, .2, .3\}$} & $\epsilon=.3$ & $\epsilon=.1$ & $\epsilon=.1$ \\
{\scshape{Metis}} & \footnotesize{$\epsilon \in \{.01, .05, .1, .2, .3\}$} & $p=.001$ & $p=.0005$ & $p=.001$ \\
\hline
\multirow{2}{*}{\scshape{Polaris}} & \footnotesize{$m \in \{1,2,3\}$} & $m=3$ & $m=2$ & $m=1$ \\ 
& \footnotesize{$v$ varies by city\tnote{*} } & $v=1k$ & $v=2k$ & $v=50k$ \\
\bottomrule
\end{tabular}
\begin{tablenotes}
\item[*]Florence $v\in \{ 1k, 15k, 50k \}$, Milan $v\in \{ 2k, 30k, 100k \}$, and Rome $v\in \{ 5k, 50k, 300k \}$.\\
\end{tablenotes}
\end{threeparttable}
\caption{Ranges of parameter values explored for each algorithm and the best values obtained for each approach.
The best parameter values for an algorithm are those leading the the lowest CO2 emissions in the city.}
\label{tab:best_params}
\end{table}

\subsection{Measures} \label{sec:measures}
We assess the effectiveness of {\scshape{Polaris}} with three measures: CO2 emissions, number of regulated intersections, and number of most popular roads.

\subsubsection*{\bf CO2 emissions}
We use SUMO \cite{Microscopic2018Lopez, cornacchia2022how, cornacchia2023} to account for vehicle interactions on the road network.  
For each city and algorithm, we generate $N$ routes and simulate their interaction within SUMO during one peak hour, uniformly selecting a route's starting time during the hour.

To estimate CO2 emissions related to the trajectories produced by the simulation, we use the HBEFA3 emission model ~\cite{infras2013handbook, bohm2022gross, cornacchia2022how}, which estimates the vehicle's instantaneous CO2 emissions at a trajectory point $j$ as:
\begin{equation}
\mathcal{E}(j) = c_0 + c_1sa + c_2sa^2 + c_3s + c_4s^2 + c_5s^3 
\end{equation}
where $s$ and $a$ are the vehicle's speed and acceleration in point $j$, respectively, and $c_0,\dots,c_5$ are parameters depending on emission and vehicle type, taken from the HBEFA database ~\cite{krajzewicz2015second}.
To obtain the total CO2 emissions, we sum the emissions corresponding to each trajectory point of all vehicles in the simulation.

\subsubsection*{\bf Regulated Intersections}
We quantify the frequency of regulated intersections encountered along a route, reflecting the route's complexity \cite{johnson2017beautiful}. 
Specifically, we categorise intersections into two types: \emph{(i)} regulated intersections, which have a traffic light or a right-before-left rule; \emph{(ii)} unregulated intersections, which have two or more incoming and/or outgoing edges.
The number of regulated intersections captures the complexity of alternative routes, measuring how smoothly travel can occur with minimal interruptions. 
A low value of this measure indicates simple routes and potentially better driver experience.

\subsubsection*{\bf Most Popular Roads} We quantify the number of popular road edges (those with high $K_{\text{road}}$) that the routes traverse. Specifically, we categorise road edges into low, medium and high popularity using an equal-sized logarithmic binning on the $K_{\text{road}}$ distribution.
The number of popular road edges encountered reflects how alternative routes tend to pass through highly trafficked roads that are already travelled by many vehicles. A low value for this measure indicates that alternative routes prefer less congested roads.

\section{Results} \label{results}

\begin{figure*}
    \centering    \includegraphics[width=0.85\textwidth]{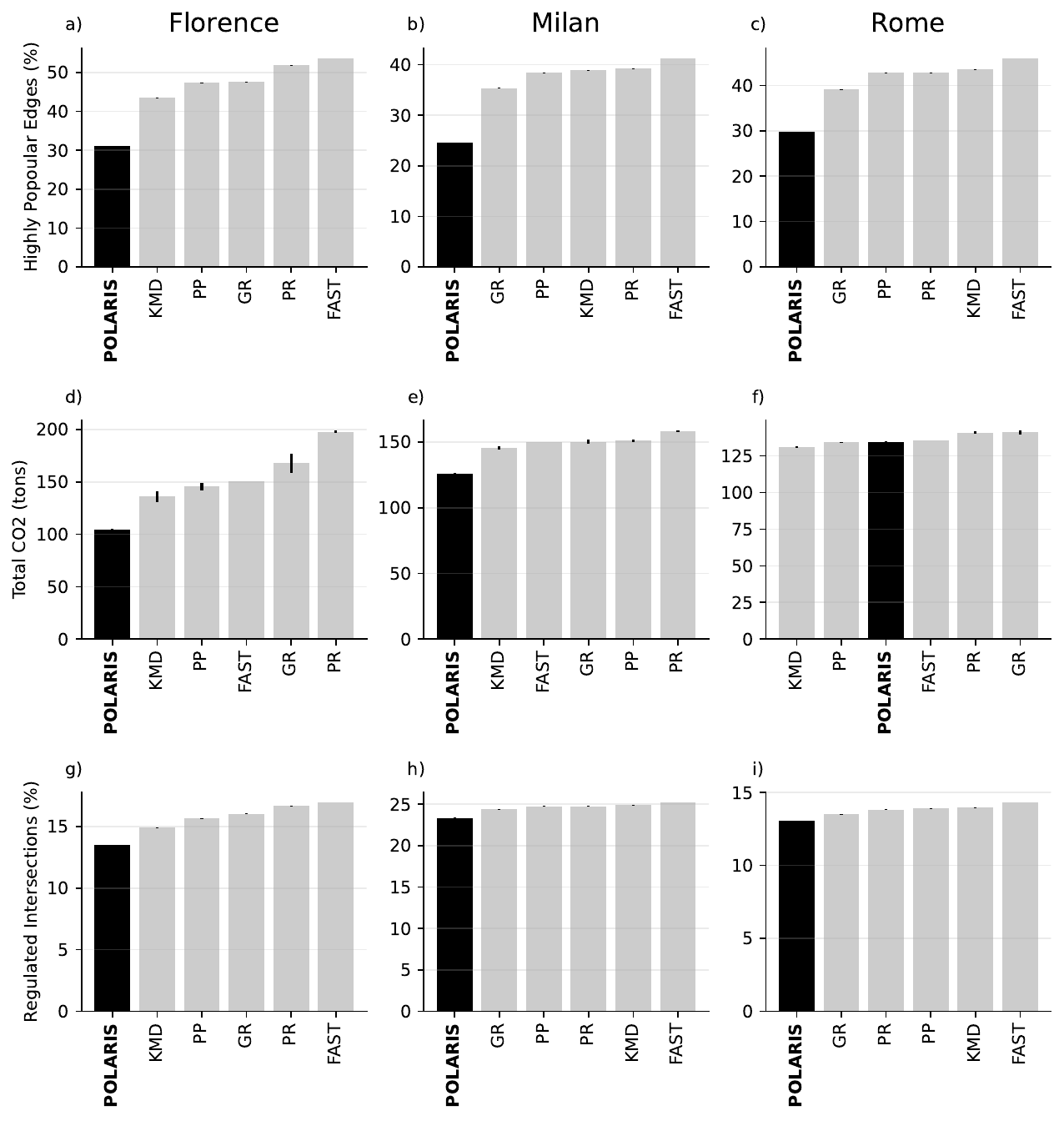}
    \caption{Comparison of {\scshape{Polaris}} (black bar) with the baselines in Florence, Milan, and Rome regarding the usage of highly popular edges (a-c, in \%), CO2 emissions (d-f, in tons), and traversed regulated intersections (g-i, in \%).
    We show the average and the standard deviation across five simulations.
    }
    \label{fig:results}
\end{figure*}

Figure \ref{fig:results} and Table \ref{tab:summary_results} compare {\scshape{Polaris}}'s performance with all baselines across all cities and measures.
We present the results regarding the combination of parameter values leading to the lowest CO2 emissions (see Table \ref{tab:summary_results}). 
The results presented refer to the average across five simulations.

As shown in Figure \ref{fig:results}a-c, the alternative routes generated by {\scshape{Polaris}} have the lowest percentage of edges with high popularity, with values of 31.07\% in Florence, 24.53\% in Milan, and 29.79\% in Rome. 
The top-performing individual baseline following {\scshape{Polaris}} uses a significantly higher percentage of edges with high popularity: 43.51\% in Florence (KMD), 35.37\% in Milan (GR), and 39.19\% in Rome (GR). 
Notably, the routes generated by the worst-performing baseline (FAST) use from 41.25\% (Milan) to 53.63\% (Florence) of popular road edges.
Across all cities, {\scshape{Polaris}} consistently uses fewer highly popular edges, demonstrating its ability to avoid congested roads and distribute the traffic more evenly.

Compared to the baselines, {\scshape{Polaris}} also exhibits lowest CO2 emissions in Florence and Milan, and the third lowest in Rome (see Figure \ref{fig:results}d-f).
In particular, when compared to KMD, {\scshape{Polaris}} achieves a CO2 reduction of 23.57\% in Florence and 13.33\% in Milan and a slight increase of 2.52\% in Rome. 

Among the individual baselines, KMD shows the lowest CO2 emissions in Florence, Milan, and Rome. 
Despite its simplicity, PP exhibits the second-lowest (Florence, Milan) and the fourth-lowest (Rome) CO2 emissions.  
In Florence and Milan, PR results in the highest CO2 emissions, whereas in Rome, GR exhibits the highest CO2 emissions.

{\scshape{Polaris}} exhibits the lowest percentage of regulated intersections encountered along alternative routes: 13.49\% in Florence, 23.35\% in Milan, and 13.04\% in Rome. 
This represents a substantial reduction 
over the best baseline algorithm in each city, which is KMD in Florence (14.93\%) and GR in Milan (24.39\%) and Rome (13.53\%).
Overall, our approach generates alternative routes that are simpler than existing approaches, minimizing the number of regulated intersections.

\begin{table*}[!ht]
\begin{tabular}{l|l c c c}
\toprule
\multicolumn{1}{c}{} & \multicolumn{1}{l}{\bf } & \textbf{\begin{tabular}[c]{@{}c@{}}Highly Popular Edges (\%)\end{tabular}} & \textbf{Total CO2 (tons)} &  \textbf{Regulated Intersections (\%)} \\ 
\midrule
& FAST & 53.63 & 150.45 & 16.93 \\
& PP & 47.39 {\footnotesize (.11)} & 145.84 {\footnotesize (3.58)} & 15.65 {\footnotesize (.05)} \\
& PR & 51.96 {\footnotesize (.04)} & 198.04 {\footnotesize (1.02)} & 16.66 {\footnotesize (.02)} \\
& GR & 47.62 {\footnotesize (.07)} & 167.91 {\footnotesize (8.91)} & 16.03 {\footnotesize (.02)} \\
& KMD & 43.51 {\footnotesize (.03)} & 136.22 {\footnotesize (5.17)} & 14.93 {\footnotesize (.02)} \\
\rowcolor{gray!10}
& {\scshape{Metis}} & 38.51 & \textbf{102.11} & 13.87 \\
\rowcolor{lavender}
\multirow{-7}{*}{\rotatebox[origin=c]{90}{Florence}} & \textbf{{\scshape{Polaris}}} & \textbf{31.07} {\footnotesize (.05)} & 104.09 {\footnotesize (1.47)} & \textbf{13.49} {\footnotesize (.01)} \\ 
\hline
& FAST & 41.25 & 149.95 & 25.16 \\
& PP & 38.41 {\footnotesize (.08)} & 150.99 {\footnotesize (.88)} & 24.74 {\footnotesize (.03)} \\
& PR & 39.22 {\footnotesize (.05)} & 158.25 {\footnotesize (.47)} & 24.75 {\footnotesize (.01)} \\
& GR & 35.37 {\footnotesize (.06)} & 150.24 {\footnotesize (1.70)} & 24.39 {\footnotesize (.04)} \\
& KMD & 38.88 {\footnotesize (.05)} & 145.55 {\footnotesize (1.06)} & 24.86 {\footnotesize (.03)} \\
\rowcolor{gray!10}
& {\scshape{Metis}} & 25.88 & 134.23 & 23.17 \\
\rowcolor{lavender}
\multirow{-7}{*}{\rotatebox[origin=c]{90}{Milan}} & \textbf{{\scshape{Polaris}}} & \textbf{24.53} {\footnotesize (.06)} & \textbf{126.15} {\footnotesize (.55)} & \textbf{23.35} {\footnotesize (.02)} \\
 \\ 
\hline
& FAST & 45.97 & 135.73 & 14.30 \\
& PP & 42.85 {\footnotesize (.06)} & 134.21 {\footnotesize (.22)} & 13.92 {\footnotesize (.01)} \\
& PR & 42.88 {\footnotesize (.02)} & 140.99 {\footnotesize (.86)} & 13.84 {\footnotesize (.02)} \\
& GR & 39.19 {\footnotesize (.06)} & 141.05 {\footnotesize (1.24)} & 13.53 {\footnotesize (.01)} \\
& KMD & 43.58 {\footnotesize (.05)} & \textbf{131.12} {\footnotesize (.50)} & 13.97 {\footnotesize (.01)} \\
\rowcolor{gray!10}
& {\scshape{Metis}} & 32.56 & 135.88 & \textbf{12.13} \\
\rowcolor{lavender}
\multirow{-7}{*}{\rotatebox[origin=c]{90}{Rome}} & \textbf{{\scshape{Polaris}}} & \textbf{29.79} {\footnotesize (.04)} & 134.43 {\footnotesize (.36)} & 13.04 {\footnotesize (.02)} \\ 
\bottomrule
\end{tabular} 
\caption{Comparison of {\scshape{Polaris}} (highlighted in blue) with baselines (KMD, GR, PR, PP, FAST) and the collective approach ({\scshape Metis}, highlighted in gray) regarding highly popular road edges, CO2 emissions, and traversed regulated intersections. 
For each algorithm, the results refer to the parameter configuration that minimises the CO2 levels in the city. The standard deviation is in parentheses for non-deterministic methods. 
The best results for each measure and city are highlighted in bold. }
\label{tab:summary_results}
\end{table*}

\begin{figure}
    \centering   \includegraphics[width=1\columnwidth]{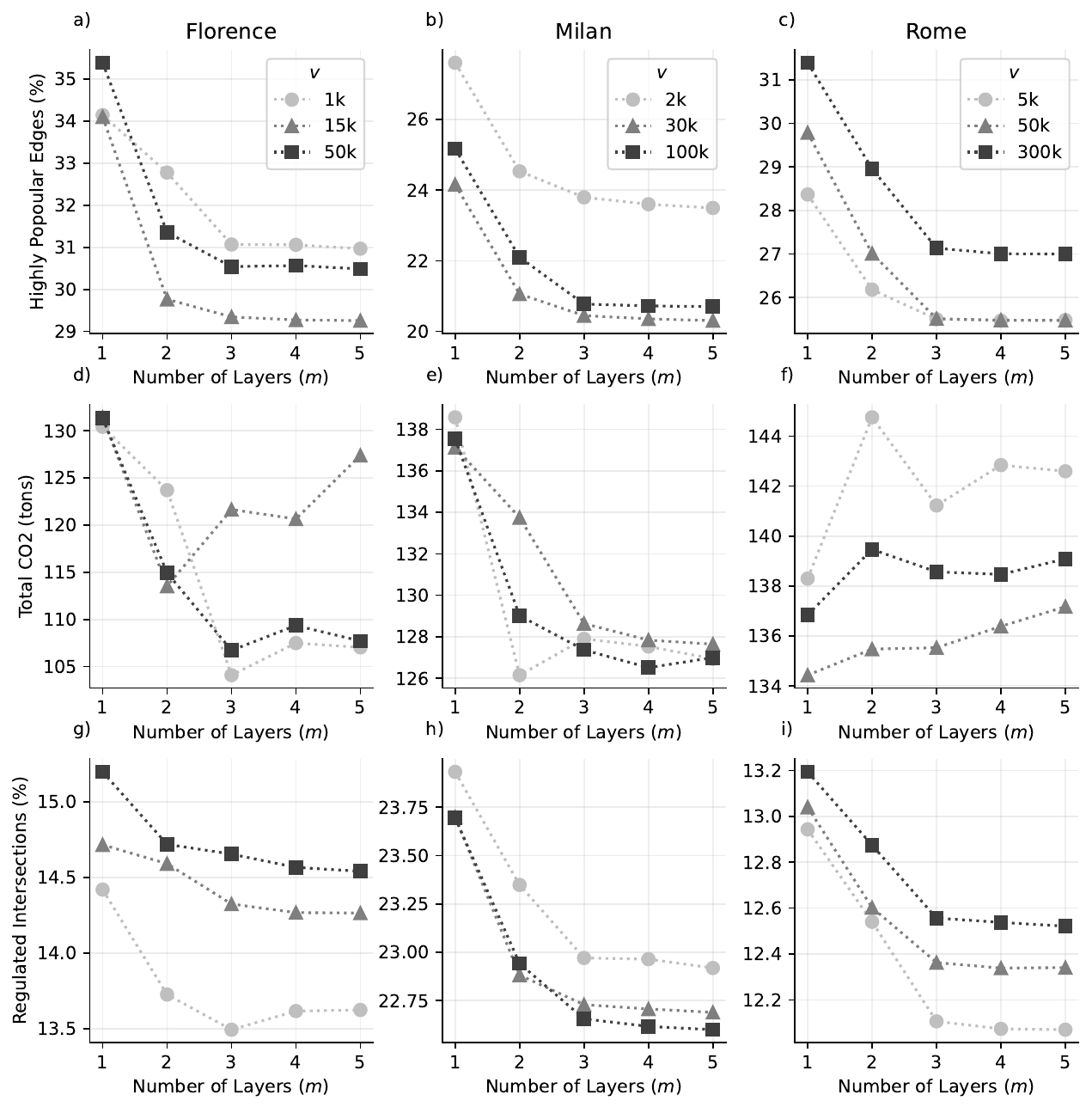}
    \caption{
     Relationship between {\scshape Polaris} parameter $m$ (number of $K_{\text road}$ layers) and percentage of highly popular edges (a-c), total CO2 emissions (d-f), and traversed regulated intersections (g-i) for three values of $v$ (number of trips used to compute $K_{\text road}$ values). }
    \label{fig:ablation}
\end{figure}

\paragraph{Parameter Sensitivity}
We analyse the relationship between the number $m$ of $K_{\text{road}}$ layers employed in {\scshape Polaris}, the number $v$ of OD pairs used to compute $K_{\text{road}}$, and three performance measures (Figure \ref{fig:ablation}).

The percentage of highly popular edges decreases noticeably from one to three $K_{\text{road}}$ layers, then stabilises. 
This trend is consistent across cities and for different numbers of OD pairs, with the most significant drop occurring between one and two $K_{\text{road}}$ layers. 

In Florence and Milan, CO2 emissions decrease significantly from one to two layers, then stabilise as more $K_{\text{road}}$ layers are employed. 
This trend is consistent across different numbers of OD pairs. In Rome, increasing the number of layers is not beneficial: using one $K_{\text{road}}$ layer in {\scshape Polaris} leads to the lowest emissions in the city. This implies that the most effective number of layers depends on the city where alternative routing is implemented.

The percentage of traversed regulated intersections also shows a decrease with the number of $K_{\text{road}}$ layers. The most significant drop is from one to two layers, with smaller declines seen afterwards. 
This trend is consistent in all the cities and for all $v$ values.

In summary, increasing the number of $K_{\text{road}}$ layers generally reduces the percentage of highly popular edges and the percentage of regulated intersections crossed. It also reduces CO2 emissions in two of three cities. These results suggest that using more $K_{\text road}$ layers in {\scshape Polaris} actually contribute to a more evenly distributed traffic load, reducing congestion on popular roads, creating simpler routes and lowering overall emissions.

\paragraph{Comparison with a coordinated approach}
We also compare {\scshape{Polaris}} against {\scshape{Metis}}, a traffic assignment approach that coordinates vehicles to diversify routes \cite{cornacchia2023}.
{\scshape{Metis}} is known to outperform existing traffic assignment approaches (including alternative routing ones) in terms of environmental impact, reducing considerably CO2 emissions \cite{cornacchia2023}. 

While {\scshape{Polaris}} generates individual routes without any driver coordination, {\scshape{Metis}} coordinates vehicles by considering the specific timing of route requests. {\scshape{Metis}} discourage a vehicle from traveling on road edges where other vehicles are expected to be when the vehicle arrives at those edges.
Despite this crucial difference, which is advantageous for {\scshape{Metis}} in distributing traffic more evenly, the two algorithms achieve comparable performance. 

{\scshape{Polaris}} uses $\approx$ 1.5-7\% fewer edges with high popularity. Furthermore, our algorithm produces only 1.94\% more CO2 emissions in Florence than {\scshape{Metis}}, and even fewer CO2 emissions in Milan (6.02\%) and Rome (1.07\%). 
Moreover, the percentage of traversed regulated intersections of {\scshape{Polaris}} are comparable to those of {\scshape{Metis}}. 

These results highlight our approach's strengths as an individual routing algorithm, achieving performance comparable to, or even better, than more sophisticated algorithms that exploit vehicle coordination.

\begin{figure}[!ht]
    \centering
    \includegraphics[width=1\linewidth]{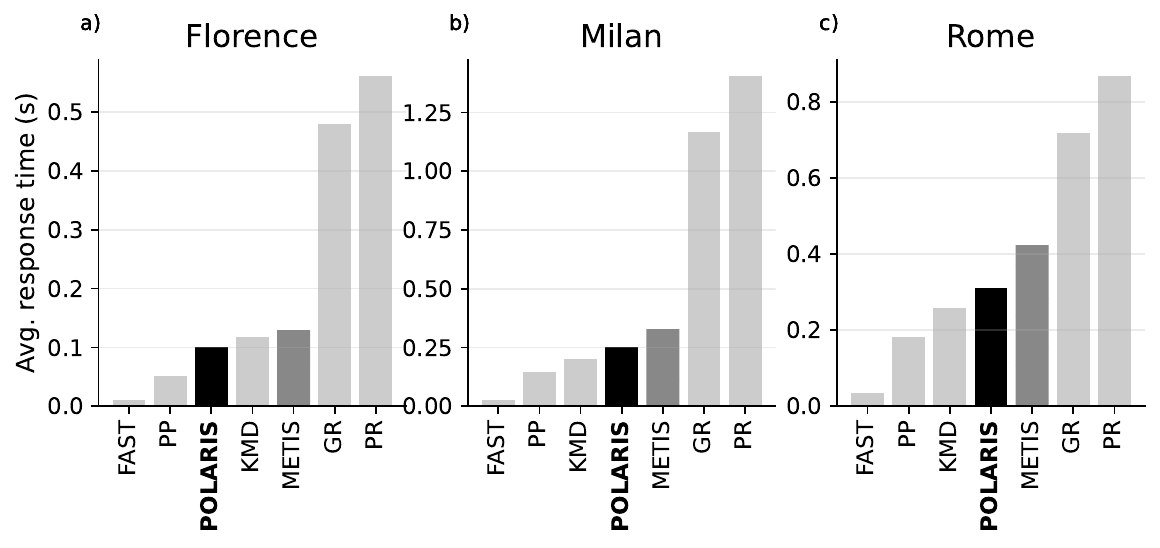}
    \caption{Comparison of the response time (in seconds) of {\scshape Polaris}, the baselines, and the coordinated approach for Florence, Milan and Rome.}
    \label{fig:execution_time}
\end{figure}

\paragraph{Execution Time}
Figure \ref{fig:execution_time} compares the average response time of {\scshape{Polaris}} and the baseline algorithms on a commodity machine with the following hardware configuration: 16 Intel(R) Core(TM) i9-9900 CPU 3.10GHz processors with 31GB RAM.

FAST is the quickest algorithm as it only requires a single shortest-path computation. 
PP is the second-fastest algorithm in every city. 
On the other hand, GR and PR are notably slow. GR needs numerous iterations to find a new distinct route since it only randomizes edge weights, often converging back to the original fastest path. PR is the most time-consuming algorithm as it requires several iterations to discover a new path by only randomizing the edge weights of the previously computed route.

{\scshape{Polaris}}'s response time is similar to PP, KMD and {\scshape Metis} but much lower than the GR and PR. In Florence, {\scshape Polaris} ranks as the third-fastest approach and fourth-fastest in Milan and Rome. 
Unlike PP, {\scshape{Polaris}} requires multiple graph weight updates, which are more expensive than updating a simple path's weight. Nevertheless, our solution results faster than GR, requiring fewer steps to compute $k$ distinct routes.
The average response time of {\scshape{Polaris}} is within the same order of magnitude of the baselines: 0.1 seconds per request in Florence, 0.25 seconds in Milan, and 0.31 seconds in Rome. This makes {\scshape{Polaris}} suitable for real-time applications where both efficiency and promptness are critical.

\section{Conclusion}

In this paper, we proposed {\scshape Polaris}, an alternative routing algorithm that minimises the use of highly popular roads. 
Our algorithm resulted in a good trade-off between CO2 emissions mitigation and smooth route options. 

In the context of the ongoing discourse on how to manage the feedback loop of human-AI interaction \cite{pedreschi2024humanai}, 

our study suggests the importance of integrating dynamic mechanisms like
ours into alternative routing algorithms to distribute traffic more evenly across the road network.

Future improvements of {\scshape{Polaris}} may involve devising a deterministic approach for selecting the random set of routes used to estimate road popularity or combining this random set with real mobility data.
Additionally, we may explore integrating route scoring mechanisms \cite{cornacchia2023} to account for real-time traffic conditions.
Route scoring would also allow further penalization based on the popularity of entire routes rather than just the popularity of single road edges.

\section*{Author Contributions}
LL conceptualized the work, developed and tested all algorithms, and made experiments;
GC conceptualized the work, preprocessed data, made experiments, made plots and figures and wrote the paper;
LP conceptualized the work, wrote the paper, and supervised the research.

\section*{Acknowledgments}
This work has been partially supported by: EU project H2020 SoBigData++ G.A. 871042; PNRR (Piano Nazionale di Ripresa e Resilienza) in the context of the research program 20224CZ5X4 PE6 PRIN 2022 “URBAI – Urban Artificial Intelligence” (CUP B53D23012770006), funded by the European Commission under the Next Generation EU programme; and by PNRR - M4C2 - Investimento 1.3, Partenariato Esteso PE00000013 - ”FAIR -- Future
Artificial Intelligence Research” -- Spoke 1 ”Human-centered AI”, funded by the European
Commission under the NextGeneration EU programme.


\bibliographystyle{ACM-Reference-Format}
\bibliography{biblio}





\end{document}